\documentclass[aps,prb,twocolumn]{revtex4}
\usepackage{epsfig,amsmath,amssymb} 
\newcommand{\be}{\begin{equation}}
\newcommand{\ee}{\end{equation}}

\newcommand{\bit}{\begin{itemize}}
\newcommand{\eit}{\end{itemize}}
\newcommand{\bea}{\begin{eqnarray}}
\newcommand{\eea}{\end{eqnarray}}
\newcommand{\Kagome}{{kagom\'e} }
\newcommand{\Neel}{N\'{e}el }
\renewcommand{\l}{\left}
\renewcommand{\r}{\right}
\usepackage{epsfig}

\begin{document}
\title
{

Quantum $J_1$--$J_2$ antiferromagnet on the stacked square lattice: 
Influence of the interlayer coupling on the ground-state magnetic ordering 
}
\author
{
D. Schmalfu{\ss},$^{1}$ R. Darradi,$^{1}$ J. Richter,$^{1}$ J.~Schulenburg,$^2$
 and D. Ihle$^{3}$ 
}
\affiliation
{
$^{1}$Institut f\"ur Theoretische Physik, Universit\"at
Magdeburg, 39016 Magdeburg, Germany \\   
$^{2}$Universit\"{a}tsrechenzentrum,
             Universit\"{a}t Magdeburg, D-39016 Magdeburg, Germany\\
$^{3}$Institut f\"ur Theoretische Physik, Universit\"at Leipzig, 04109 Leipzig,
Germany\\
}

\begin{abstract}
Using the coupled-cluster method (CCM)
and the rotation-invariant Green's function method (RGM), 
we study the influence of the interlayer coupling $J_\perp$ 
on  the
magnetic  ordering in the ground state of the 
spin-$1/2$ $J_1$-$J_2$ frustrated Heisenberg antiferromagnet 
($J_1$-$J_2$ model)
on the stacked  square lattice.  
In agreement with known results 
 for the $J_1$-$J_2$ model on the strictly two-dimensional square lattice
($J_\perp=0$)
we find that
the phases with magnetic
long-range order at small $J_2< J_{c_1}$ and large $J_2> J_{c_2}$ are
separated by a  magnetically disordered (quantum paramagnetic) 
ground-state phase.
Increasing the interlayer coupling $J_\perp>0$ 
the parameter region of this phase decreases,
and, finally, the quantum paramagnetic 
phase disappears 
for quite small  $J_\perp \sim
0.2\; - \; 0.3 J_1$.
\end{abstract}

\pacs{Valid PACS appear here}
\maketitle


The properties of the frustrated spin-$1/2$ Heisenberg
antiferromagnet (HAFM) with nearest-neighbor $J_1$ and competing
next-nearest-neighbor $J_2$ coupling 
($J_1$-$J_2$ model) on the square lattice have attracted a
great deal of interest during the last fifteen years (see, e.g., Refs.
\onlinecite{chandra88,dagotto89,schulz,richter,bishop98,
singh99,sushkov01,capriotti00,capriotti01,siu01,singh03,roscilde04}
and references therein). The recent synthesis of layered magnetic materials
\cite{melzi00,rosner02} which can be described by the $J_1$-$J_2$ 
model has stimulated a renewed interest in this model.
It is well-accepted that
the model exhibits two magnetically long-range ordered phases 
at small and at  large $J_2$ 
separated by an intermediate  quantum paramagnetic phase 
without magnetic long-range order
(LRO)
in the parameter region 
$J_{c_1} < 
J_2 < J_{c_2}$, where $J_{c_1} \approx 0.4$
and  $ J_{c_2} \approx 0.6$.
The ground state (GS) at low $J_2 < J_{c_1}$ 
exhibits semi-classical N\'eel magnetic LRO with the
magnetic wave vector  
  ${\bf Q}_{0}=(\pi ,\pi )$.
The GS at large $J_2 > J_{c_2} $ shows so-called 
collinear magnetic LRO with  the 
magnetic wave vectors ${\bf Q}_{1} =(\pi , 0)$
or ${\bf Q}_{2} =(0 ,\pi )$. 
These two collinear states are characterized by a parallel spin orientation of 
nearest neighbors in vertical (horizontal) direction and an antiparallel 
spin orientation of  nearest neighbors in horizontal (vertical) 
direction.
The properties of the  intermediate quantum paramagnetic phase are  
still under discussion, however, a valence-bond crystal phase 
seems to be most favorable.\cite{dagotto89,schulz,richter,capriotti00,sushkov01}

The properties of quantum magnets strongly depend on the
dimensionality.\cite{lnp04}
Though the tendency to order is
more pronounced in three-dimensional (3d)
systems than in low-dimensional ones, a magnetically disordered
phase can also be
observed in frustrated 3d systems such as the HAFM on
the pyrochlore lattice \cite{canals98} or on 
the stacked \Kagome lattice.\cite{schmal04}
On the other hand, recently it has been found that the  3d 
$J_1$-$J_2$ model on the body-centered cubic lattice does not have an 
intermediate quantum paramagnetic phase.\cite{schmidt02,oitmaa04}
Moreover, in experimental realizations of the $J_1$-$J_2$ model
the magnetic couplings are expected to be not strictly 2d, 
but a finite interlayer coupling $J_\perp$ is present.
For example, recently Rosner et al.\cite{rosner02}
have found  $J_\perp/J_1 \sim 0.07$  for Li$_2$VOSiO$_4$, a 
material which can be described by a square lattice $J_1$-$J_2$ model 
with large $J_2$.\cite{melzi00,rosner02} 

This motivates us to consider  an extension
of the $J_1$-$J_2$ model, namely the $J_1$-$J_2$ spin-$1/2$ HAFM 
on the stacked square lattice described by the Hamiltonian
\begin{eqnarray}
\label{ham}
H&=&\sum_n\Bigg(J_1\sum_{\langle ij \rangle}{\bf s}_{i,n} \cdot {\bf s}_{j,n}
+J_2 \sum_{[ ij ]}{\bf s}_{i,n} \cdot {\bf s}_{j,n}\Bigg) \nonumber\\
&& + J_\perp \sum_{i,n} {\bf s}_{i,n} \cdot {\bf s}_{i,n+1},
\end{eqnarray}
where $n$ labels the layers 
and  $J_\perp \ge 0$ is the interlayer coupling.
The expression in brackets
represents the 
 $J_1$-$J_2$ model of the layer $n$ with intralayer couplings
$J_1=1$ 
and $J_2 \ge 0$.
The main problem we would like to study is the influence of 
$J_\perp$ on the existence of the intermediate quantum paramagnetic
GS phase. Note that the exact diagonalization 
widely 
used for the study of the strictly 2d $J_1$-$J_2$ model, 
see, e.g., Refs.
\onlinecite{dagotto89,schulz,richter}, is not appropriate for the
3d problem under consideration.
Therefore, we use   
the coupled-cluster method
(CCM)\cite{bishop98,bishop91,zeng98,krueger00,rachid05,farnell05} and 
the  rotation-invariant Green's function method
(RGM).\cite{siu01,schmal04,kondo72,bern02,siu00,schmal05} 
Both methods have been successfully applied to quantum spin
systems in arbitrary dimension and are able to deal with frustration.  

Let us briefly illustrate some basic features of the CCM. 
For more details the reader is referred to Refs. 
\onlinecite{bishop98} and \onlinecite{bishop91,zeng98,krueger00,rachid05,farnell05}.   
The starting point for the CCM calculation is the choice of 
a reference state 
$|\Phi\rangle$.  
For $|\Phi\rangle$ of the considered spin system we choose the
two-sublattice \Neel state for small $J_2$ but a  collinear state 
for large $J_2$.
To treat each site equivalently we perform a rotation of the local axis of
the spins such that all spins in the reference state align along 
the negative $z$ axis, i.e., in the rotated coordinate frame
we have $|\Phi\rangle \hspace{-3pt} = 
\hspace{-3pt}|\hspace{-3pt}\downarrow \rangle|\hspace{-3pt}\downarrow 
\rangle |\hspace{-3pt}\downarrow \rangle \ldots \,\,$. 
Note that in this new frame the   Hamiltonian 
is modified, 
and 
$|\Phi\rangle \hspace{-3pt} = 
\hspace{-3pt}|\hspace{-3pt}\downarrow \rangle|\hspace{-3pt}\downarrow 
\rangle |\hspace{-3pt}\downarrow \rangle \ldots \,\,$ is  not an eigenstate  of this modified 
Hamiltonian, 
 see, e.g. Refs. \onlinecite{bishop91,zeng98,rachid05}. 
For the ket GS   $|\Psi\rangle$ with
$H|\Psi\rangle = E|\Psi\rangle$
an exponential ansatz
$|\Psi\rangle = e^S|\Phi\rangle$ is used, where 
the correlation operator $S$ is given by $S = \sum_{I \neq
0}{\cal S}_IC_I^+$.
The $C_I^+$ represent a set of multi-spin creation operators $C_I^+=s_i^+ \; , \;
s_i^+ s_j^+ \; , \; s_i^+ s_j^+ s_k^+\; , \; \ldots \;$. 
The application of all the $C_I^+$ on $|\Phi\rangle$ creates a complete set of 
states, which may contribute to $|\Psi\rangle$. 
The correlation operator $S$ contains the  
coefficients ${\cal S}_I$ which are
determined by  requiring that the expectation value of $H$
is a minimum. 
The order parameter $M$ is given
by the expectation value of $s_i^z$.

For the considered   quantum many-body model
it is necessary to use approximations 
in order to truncate the expansion 
of $S$.
We use the well elaborated  LSUB$n$ scheme\cite{bishop91,zeng98,rachid05} 
in which in the correlation operator $S$
all multi-spin correlations over all distinct locales on the lattice defined by $n$ or fewer
contiguous sites are taken into account. 
For example, within the 
LSUB4 scheme one includes multi-spin creation operators of one, two, three or 
four spins distributed on 
arbitrary clusters of four contiguous lattice sites. 
The number of these fundamental configurations can be reduced exploiting lattice
symmetry and conservation laws. In the CCM-LSUB8 approximation we have 
finally 25953 (43070) fundamental configurations 
for the \Neel (collinear) reference state.  
To solve the set of the corresponding ket 
equations we use parallel computing.\cite{cccm} 

Since the LSUB$n$ 
approximation becomes exact for 
$n \to \infty$, it 
is useful to extrapolate the 'raw' LSUB$n$
data to $n \to \infty$. 
An appropriate extrapolation rule for the order parameter of 
systems showing a GS order-disorder transition
is the 'leading power-law' extrapolation\cite{rachid05} 
$ M(n)=c_0+c_1(1/n)^{c_2} ,
$
where the results  of the LSUB4,6,8 approximations are used for the 
extrapolation.
For the GS energy per spin 
$
e(n) = a_0 + a_1(1/n^2) + a_2(1/n^4)$
is a reasonable extrapolation ansatz.\cite{krueger00}

Next we give a brief illustration
of the spin-rotation-invariant Green's function method.\cite{kondo72,winter97}
More details can be found in
Refs. \onlinecite{siu00,siu01,schmal04} and \onlinecite{schmal05}.   
Considering the
equations of motion for the commutator Green's function 
$\left\langle\left\langle
s^{+}_{{\bf{q}}};s^{-}_{{\bf{q}}}\right\rangle\right\rangle_{\omega}$
and supposing spin-rotation invariance, i.e.
$\langle s^{z}_{m}\rangle\equiv 0$, we get
$
\omega^{2}\left\langle\left\langle
s^{+}_{{\bf{q}}};s^{-}_{{\bf{q}}}\right\rangle\right\rangle_{\omega}
=\big\langle\big[i\dot{s}^{+}_{{\bf{q}}},s^{-}_{{\bf{q}}}\big]_{-}
\big\rangle+\left\langle\left\langle
-\ddot{s}^{+}_{{\bf{q}}};s^{-}_{{\bf{q}}}\right\rangle\right\rangle_{\omega}
$.
To
treat the operator $\ddot{s}^{+}_{{\bf{q}}}$ containing products of three 
spin operators along nearest-neighbor sequences, a decoupling procedure in
the spirit of Ref.~\onlinecite{kondo72} is performed. 
For example,
the operator product $s^{-}_{A}s^{+}_{B}s^{+}_{C}$ is replaced by
$\eta_{A,B}\l\langle s^{-}_{A}s^{+}_{B}\r\rangle
s^{+}_{C}+\eta_{A,C}\l\langle s^{-}_{A}s^{+}_{C}\r\rangle s^{+}_{B}$, where
$A,B,C$ represent spin sites. The introduction of vertex parameters
$\eta_{\gamma,\mu}$ is aimed to improve the approximation and to fulfill
fundamental constraints like the sum rule. 

By analogy with Refs.~\onlinecite{siu01} and \onlinecite{siu00} we use
four different vertex parameters, namely
$\eta_{1\parallel}$ related to the correlator $c_{1,0,0}$,
$\eta_{1\perp}$ related to $c_{0,0,1}$,
$\eta_{2}$ commonly related to  $c_{2,0,0}$,
$c_{2,1,0}$, $c_{2,2,0}$, $c_{1,0,1}$, $c_{1,1,1}$, $c_{0,0,2}$,
and 
$\eta_{3}$ related to  $c_{1,1,0}$.
The
correlators are defined as $c_{k,l,m}\equiv c_{{\bf{R}}}=\l\langle
s^{+}_{{\bf{0}}}s^{-}_{{\bf{R}}}\r\rangle=2\l\langle
{\bf{s}}_{{\bf{0}}}{\bf{s}}_{{\bf{R}}}\r\rangle/3$ with the lattice vector
${\bf{R}}=k{\bf{a}}_{1}+l{\bf{a}}_{2}+m{\bf{a}}_{3}$ 
and have to be determined selfconsistently.
Performing the  approximations mentioned above we obtain
$
\left\langle\left\langle
s^{+}_{{\bf{q}}};s^{-}_{{\bf{q}}}\right\rangle\right\rangle_{\omega}
=m_{{\bf{q}}}/(\omega^{2}-\omega^{2}_{{\bf{q}}})
$,
where for 
$m_{{\bf{q}}}$  and
$\omega^{2}_{{\bf{q}}}$
explicit equations can be given.
The equation for $\omega^{2}_{{\bf{q}}}$ contains 
the four vertex parameters and  
the nine correlators mentioned above.
The correlators can be expressed by the Green's function using
 the spectral theorem. 
To determine the four vertex parameters 
we use the sum rule 
$c_{0,0,0}=1/2$ 
and require that the static susceptibility
$\chi^{+-}_{{\bf{q}}}=-\lim_{\omega\to 0}
\left\langle\left\langle
s^{+}_{{\bf{q}}};s^{-}_{{\bf{q}}}\right\rangle\right\rangle_{\omega}$
has to be isotropic in the limit
${\bf{q}}\to{\bf{0}}$.\cite{siu00,schmal04,schmal05}
The remaining two equations
are obtained as follows:
First we use the relation
$
\eta_{3}=
(\eta_{2}e^{-J_{2}}+J_{2}\eta_{1{\parallel}})(1+J_{2})^{-1}$
which was successfully applied  in Ref.~\onlinecite{siu01} 
to the 2d $J_1$-$J_2$ model. 
This relation interpolates between 
the two limiting cases $J_2 \to 0$ and $J_2 \to \infty$ and
takes care of the relation $\lim_{J_2 \to 0}c_{1,0,0}=
\lim_{J_2 \to \infty}c_{1,1,0}$.   
Finally we use, following Ref.~\onlinecite{siu01},
an approximative  expression for the GS energy per spin
$e^{input}_0= 3J_1 c_{1,0,0} + 3J_2c_{1,1,0} + 3J_\perp
c_{0,0,1}/2 $  as  an additional input.
For the
stacked HAFM considered  we make the ansatz 
$e^{ input}_0(J_2,J_\perp)=f_1(J_\perp)+f_2(J_2)$ (note that $J_1=1$).
To fix $f_2$ 
we  use the  
exact diagonalization result for the GS energy  of the
 finite 2d $J_1$-$J_2$ model ($J_\perp=0$) of $N=32$
spins, i.e. we set $f_2(J_2)= e^{N=32}_0(J_2,J_\perp=0)$.
To fix $f_1$
we use the GS energy 
of the unfrustrated stacked square lattice $e^{SW}_0(J_2=0,J_{\perp})$
calculated by linear spin-wave theory
 and set  $f_1(J_\perp)=e^{SW}_0(J_2=0,J_{\perp})-
e^{N=32}_0(J_2=0,J_\perp=0)$
this way taking into account 
 the effect of the
interlayer coupling and a finite-size correction.

To discuss GS magnetic order-disorder transitions we consider the
magnetic order parameter.
In the RGM scheme\cite{kondo72,winter97,siu00,schmal05} 
the correlation function
$\l\langle {\bf{s}}_{{\bf{0}}}{\bf{s}}_{{\bf{R}}}\r\rangle$ at $T=0$ 
is given by
\be
\l\langle {\bf{s}}_{{\bf{0}}}{\bf{s}}_{{\bf{R}}}\r\rangle =\frac
{3}{2N}\sum_{{\bf{q}}\neq{\bf{Q}}_{j}}\frac
{m_{{\bf{q}}}}{2\omega_{{\bf{q}}}}e^{-i{\bf{q}}{\bf{R}}}
+\frac
{3}{2}\sum_{{\bf{Q}}_{j}}C_{{\bf{Q}}_{j}}e^{-i{\bf{Q}}_{j}{\bf{R}}}.
\label{eq6}
\ee
The second term (condensation part) describes LRO,
where the sum runs over 
different nonequivalent magnetic wave
vectors ${\bf{Q}}_{j}$ taking into account the possibility to have
degenerate GSs.
For model (\ref{ham}) we have ${\bf{Q}}_{0}=\l(\pi,\pi,\pi\r)$ 
for the \Neel phase and 
${\bf{Q}}_{1}=\l(\pi,0,\pi\r)$ or ${\bf{Q}}_{2}=\l(0,\pi,\pi\r)$ 
for the collinear phase.
Magnetic LRO is accompanied by a diverging
static susceptibility $\chi^{+-}_{{\bf{q}}}$ at ${\bf{q}}={\bf{Q}}$ giving
an additional equation for $C_{{\bf{Q}}}$. Note that for the collinear
phase both condensation terms are equal, i.e. $C_{{\bf{Q}}_{1}}=
C_{{\bf{Q}}_{2}}$.
The order parameter $M$ 
can be
calculated by $M^{2}=3\l|C_{{\bf{Q}}}\r|/2$.
That way, the order parameter is linked to the long-range behavior of the
correlation functions because $M$ is nonzero if 
$\lim_{\l|{\bf{R}}\r|\to\infty}\l\langle {\bf{s}}_{{\bf{0}}}{\bf{s}}_{{\bf{R}}}\r\rangle$
remains finite.

\begin{figure}
\begin{center}
\epsfig{file=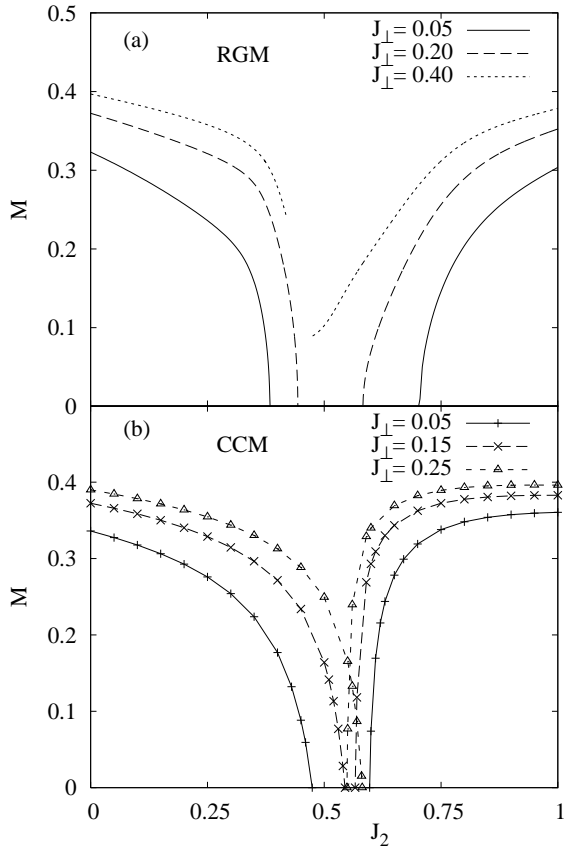,scale=0.7,angle=0.0}
\end{center}
\caption{ Magnetic order parameter $M$ versus $J_2$ for various strengths of 
the interlayer coupling $J_\perp$. (a): RGM, (b): CCM
}
\label{fig1}
\end{figure}

\begin{figure}
\begin{center}
\epsfig{file=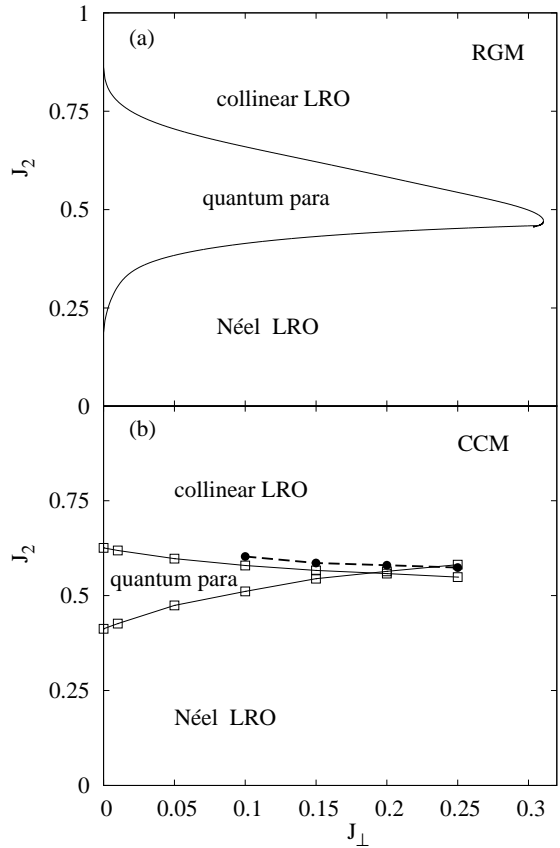,scale=0.7,angle=0.0}
\end{center}
\caption{ Ground-state phase diagram.
(a): RGM, (b): CCM.\\
The solid lines show those values of $J_2$ where the order
parameters vanish. The dashed line in (b) represents those values 
of $J_2$
where the two energies calculated for the \Neel and collinear reference state
become equal. 
}
\label{fig2}
\end{figure}

As in the 2d case the GS of the stacked model
is characterized by two  magnetically long-range ordered phases, namely
 a N\'{e}el phase
for small $J_{2}$ and a  collinear phase for
large  $J_{2}$. For not too large  
$J_\perp$ 
both magnetic phases are separated by a magnetically disordered 
quantum paramagnetic
phase, where the phase transition points 
are functions of $J_{\perp}$. 
To determine these transition points we calculate the order parameters 
for various $J_\perp$ 
to find those values $J_2=\alpha_{N{e}el}(J_\perp)$ and 
$J_2=\alpha_{coll}(J_\perp)$ where the order parameters vanish.
In Fig.~\ref{fig1} we present some typical curves showing the order parameters 
versus  $J_2$ for some values of $J_\perp$. Both approaches lead to
qualitatively comparable results. The magnetic order
parameters of both magnetically long-range ordered phases 
vanish continuously as it is typical for second-order
transitions. Note, however, that there are arguments\cite{schulz,sushkov01} 
that 
the transition from the collinear phase to the quantum paramagnetic phase
should be of first order.  
The order parameters 
are monotonously increasing with $J_\perp$, and the transition points 
$\alpha_{N{e}el}$ and $\alpha_{coll}$ move together.
In Fig.\ref{fig2} we present these
transition points in dependence on $J_\perp$.
Close to the strictly 2d case, i.e. for small $J_\perp \ll 1$, the
influence of the interlayer coupling is largest.
For a characteristic value of $J^*_\perp \approx 0.31$ $(0.19)$ for the 
RGM (CCM)
approach the transition points $\alpha_{N{e}el}$ and $\alpha_{coll}$ 
meet each other.

 For larger $J_\perp$ exceeding $J^*_\perp$ we have  a direct 
first-order transition between both types of magnetic LRO as 
it is also observed in the
classical model and in the 3d quantum $J_1$-$J_2$ model on the  
body-centered cubic lattice.\cite{schmidt02,oitmaa04}
However, the description of this first-order transition is not possible
within the RGM approach. The reason is that  the  approximative  
expression for the  GS energy per spin
$e^{input}_0$ used as an input is a smooth function of $J_2$, whereas a
first-order
GS transition is characterized by a kink in $e_0$. 
As a consequence we find that there is no solution of the system of
coupled RGM equations for parameter values being close to a first-order
transition, i.e. for $J_2 \approx 0.5$ and $J_\perp > J^*_\perp$. The order
parameter curve for $J_\perp = 0.4$ depicted in Fig.~\ref{fig1}(a) indeed shows
a small region slightly below $J_2 =0.5$, where no solution exists.  

\begin{figure}
\begin{center}
\epsfig{file=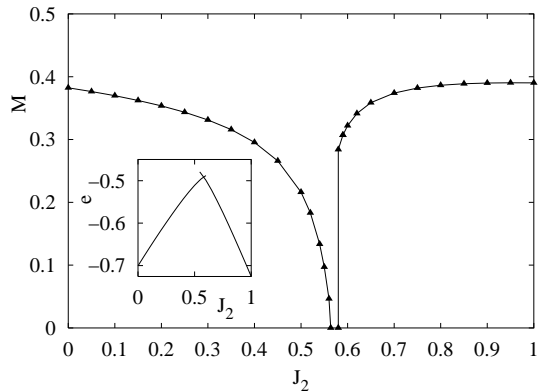,scale=0.6,angle=0.0}
\end{center}
\caption{CCM results for the energy per spin $e$ 
for both reference states (inset) and the order parameter $M$ for $J_\perp=0.2$. 
Both quantities are 
obtained by extrapolation of the 'raw' LSUB$n$
results to the limit $n \to \infty$  as explained in the text.
The energies calculated with \Neel  
and collinear reference 
states become equal  at $J_2\approx 0.58$ indicating
a first-order transition. 
For the order parameter $M$ we take that value calculated with  
the reference state of  lower CCM energy.  
}
\label{fig3}
\end{figure}

In contrast to the RGM  the CCM approach starts with two different reference
states (\Neel and collinear)
related to the two types of magnetic LRO.
Though we start our CCM calculation with
a reference state corresponding to semiclassical order, one can compute
the GS energy also in parameter regions
where semiclassical magnetic LRO is destroyed,
and  it is 
known \cite{bishop98,krueger00,rachid05,farnell05} that
the CCM yields 
precise results for the GS energy 
beyond the transition from the semiclassical magnetic phase to the
quantum paramagnetic phase. 
The necessary condition for the convergence of the CCM equations is
a sufficient overlap between the reference state and the true GS.
Hence we can add to the above discussion of
the order parameters
a comparison of the energies.
Provided that the CCM equations converge for the \Neel and the collinear reference
state far enough beyond those points where the order parameters vanish 
we can 
determine the point where both energies become equal. For the considered LSUBn 
approximations this happens for $J_\perp \gtrsim 0.1$.
In the inset of Fig.~\ref{fig3} we show the energies versus $J_2$ 
for $J_\perp =0.2$
caclulated by extrapolation.
 The corresponding 
points $J_2=\alpha'_{coll}(J_\perp)$ where
both energies meet are shown in Fig.~\ref{fig2} as dashed line.  

We obtain that both transition points 
$\alpha_{coll}$ and $\alpha'_{coll}$ are close to each other and show a
similar dependence on $J_\perp$.
Secondly, we find that at least for $J_\perp \gtrsim 0.1$ 
the energy obtained with the \Neel reference state is lower than 
that obtained with the
collinear reference state even for $J_2$ values where 
the \Neel order parameter is already zero but
the collinear order parameter is still finite. 
Thus, this energetic consideration leads to the following sequence of
zero-temperature transitions: Second-order transition from \Neel LRO to
a quantum paramagnetic phase at $J_2=\alpha_{N{e}el}$ 
and then a first-order
transition from the quantum paramagnetic phase to collinear LRO at
$J_2=\alpha'_{coll} > \alpha_{coll} > \alpha_{N{e}el}$. This behavior is illustrated 
in Fig.~\ref{fig3}, where the 
order parameter $M$ is shown versus $J_2$ for fixed $J_\perp=0.2$.  
For a certain value of $J_\perp \approx 0.23$ both transition points 
$\alpha_{N{e}el}$ and $\alpha'_{coll}$ become
equal, and one has a direct first-order transition between the two
semiclassically long-range ordered phases.

{\it   Acknowledgment:}
This work was supported by the DFG (Ri615/12-1, Ih13/7-1).


\end{document}